\RequirePackage{ifpdf}
\ifpdf 
\documentclass[pdftex]{sigma}
\else
\documentclass{sigma}
\fi

\begin{document}
\allowdisplaybreaks

\renewcommand{\PaperNumber}{019}

\FirstPageHeading

\ShortArticleName{Eigenvectors of Open Bazhanov--Stroganov Quantum
Chain}

\ArticleName{Eigenvectors of Open Bazhanov--Stroganov\\ Quantum
Chain}

\Author{Nikolai IORGOV} \AuthorNameForHeading{N. Iorgov}

\Address{Bogolyubov Institute for Theoretical Physics, 14b
Metrolohichna Str.,  Kyiv, 03143 Ukraine}
\Email{\href{mailto:iorgov@bitp.kiev.ua}{iorgov@bitp.kiev.ua}}

\ArticleDates{Received November 29, 2005, in f\/inal form January
30, 2006; Published online February 04, 2006}

\Abstract{In this contribution we give an explicit formula for the
eigenvectors of Hamiltonians of open Bazhanov--Stroganov quantum
chain. The Hamiltonians of this quantum chain is def\/ined by the
generation polynomial $A_n(\lambda)$ which is upper-left matrix
element of monodromy matrix built from the cyclic
$L$-operators. The formulas for the eigenvectors are derived using
iterative procedure by Kharchev and Lebedev and given in terms of
$w_p(s)$-function which is a root of unity analogue of
$\Gamma_q$-function.}

\Keywords{quantum integrable systems; Bazhanov--Stroganov quantum
chain}

\Classification{81R12; 81R50}

\section{Introduction}

In the papers \cite{BS, Kor} it was observed
that the six-vertex $R$-matrix at root of unity intertwines not
only the six-vertex $L$-operators, but also some other
$L$-operators (which are called cyclic $L$-operators).
These $L$-operators def\/ine $n$-particle
Bazhanov--Stroganov quantum chain (BSQC) by the standard procedure
in quantum inverse scattering method: the product of $L$-operators
def\/ines the monodromy matrix
\begin{gather} \label{mmintro}
{T}_n(\lambda)=L_1(\lambda)L_2(\lambda)\cdots L_n(\lambda)= \left(
\begin{array}{cc}
A_n(\lambda)& B_n(\lambda)\\[2mm]
C_n(\lambda)& D_n(\lambda) \end{array} \right),
\end{gather}
which in turn provides us with commuting set of operators ---
Hamiltonians of quantum chain. It was observed by Baxter that the
so-called ``Inverse SOS'' model discovered by him \cite{BaxtInv}
is equivalent to Bazhanov--Stroganov quantum chain. Thus the same
model (which is called the $\tau_2$-model) has two formulations:
the formulation as face model by Baxter and the formulation as
quantum chain (or vertex model) by Bazhanov and Stroganov. The
connection between six-vertex model, $\tau_2$-model and chiral
Potts model gave a possibility to formulate the system of
functional relations \cite{BS,BBP} for transfer-matrices of these
models. This system provides the main tool for derivation the free
energy \cite{BaxtEV} and the order parameter \cite{BaxtOP} for
chiral Potts model.

The goal of this contribution is to f\/ind common eigenvectors of
the set of commuting Hamiltomians ${\boldsymbol H}_k$,
$k=1,2,\ldots,n$, of open $n$-particle BSQC. These Hamiltonians
are def\/ined by the coef\/f\/icients of the $A_n(\lambda)$ given
by (\ref{mmintro}):
\[
A_n(\lambda)=1+\lambda {\boldsymbol H}_1+\lambda^2 {\boldsymbol
H}_2+ \cdots+\lambda^n {\boldsymbol H}_n.
\]
The main idea how to f\/ind the eigenvectors is to use iterative
procedure. Namely we build the eigenvectors of $A_n(\lambda)$
using the eigenvectors of $A_{n-1}(\lambda)$ which is the
generation function of Hamiltonians for open $(n-1)$-particle
BSQC. This procedure in essential is an adaptation of iterative
procedure by Kharchev and Lebedev \cite{KhL} for quantum Toda
chain. The mentioned idea has origin in the paper by Sklyanin
\cite{Skl}, where he used separated variables of subsystems to
construct the separated variables of the whole system. The models
where the iterative procedure was realized are relativistic Toda
chain \cite{KhLS}, Toda chain with boundary interaction \cite{IS},
periodic Bazhanov--Stroganov model \cite{GIPS}. It is worth to
note that these models do not admit algebraic Bethe ansatz
procedure because in the case of generic parameters these models
do not possess ``highest weight vectors''. The method used in this
contribution is an evolution of the method of separated variables
(or functional Bethe ansatz method).

At the end of the introduction we would like to mention that the
Bazhanov--Stroganov model at special values of parameters reduces
to the relativistic Toda chain at a root of unity. Also it is worth
noting {\it direct} (not through the chiral Potts model!) relation
\cite{BIS} between lattice formulation of the model at $N=2$ and Ising
model. This relation gave a possibility to find the eigenvalues \cite{BIS}
and the eigenvectors \cite{Lis}
of the transfer-matrix by means of auxiliary grassmann field technique.

\section[Bazhanov-Stroganov quantum chain]{Bazhanov--Stroganov quantum chain}

Let $\omega=e^{2\pi{\rm i}/N}$, $N\ge 2$. For each particle $k$,
$k=1,2,\ldots,n$, of Bazhanov--Stroganov quantum chain (BSQC) with
$n$ particles there corresponds $N$-dimensional linear space
(quantum space)~${\cal V}_k$ with the basis
$|\gamma\rangle_k$,$\gamma\in \mathbb{Z}_N$, and a pair of
operators $\{{\boldsymbol u}_k,{\boldsymbol v}_k\}$ acting on
${\cal V}_k$ by the formulas:
\begin{gather}\label{uvact}
{\boldsymbol v}_k|\gamma\rangle_k=\omega^\gamma
|\gamma\rangle_k,\qquad {\boldsymbol u}_k|\gamma\rangle_k=
|\gamma-1\rangle_k.
\end{gather}
The space of quantum states of BSQC with $n$ particles is ${\cal
V}={\cal V}_1\otimes{\cal V}_2\otimes\cdots\otimes{\cal V}_n$. We
extend the action of operators $\{{\boldsymbol u}_k,{\boldsymbol
v}_k\}$ to ${\cal V}$ def\/ining this action to be identical on
${\cal V}_s$, $s\ne k$. Thus we have the following commutation
relations
\begin{gather*}
{\boldsymbol u}_j {\boldsymbol u}_k={\boldsymbol u}_k {\boldsymbol
u}_j,\qquad {\boldsymbol v}_j {\boldsymbol v}_k={\boldsymbol v}_k
{\boldsymbol v}_j,\qquad {\boldsymbol u}_j {\boldsymbol v}_k=
\omega^{\delta_{j,k}} {\boldsymbol v}_k{\boldsymbol u}_j.
\end{gather*}

For each particle of BSQC model we put into correspondence the
cyclic $L$-operator
\begin{gather}\label{bazh_strog}
L_k(\lambda)=\left(
\begin{array}{cc}
1+\lambda \varkappa_k {\boldsymbol v}_k &  \lambda {\boldsymbol u}_k^{-1} (a_k-b_k {\boldsymbol v}_k)\\
{\boldsymbol u}_k (c_k-d_k {\boldsymbol v}_k) & \lambda a_k c_k +
{\boldsymbol v}_k {b_k d_k}/{\varkappa_k }
\end{array}
\right),\qquad k=1,2,\ldots,n,
\end{gather}
where $\{a_k,b_k,c_k,d_k,\varkappa_k\}$ are (in general complex)
parameters attached to $k$th particle. In the papers \cite{BS,Kor} it
was observed that the six-vertex $R$-matrix
\begin{gather*}
{R}(\lambda,\mu)=\left(\begin{array}{cccc}
\lambda-\omega\mu & 0 & 0 & 0 \\[1mm]
0 & \omega(\lambda-\mu) & \lambda (1-\omega) & 0 \\[1mm]
0 & \mu (1-\omega) & \lambda-\mu & 0 \\[1mm]
0 & 0 & 0 & \lambda-\omega\mu \end{array}\right).
\end{gather*}
at root of unity $\omega=e^{2\pi {\rm i}/N}$ intertwines not only
the six-vertex $L$-operator, but also the cyclic
$L$-operators~(\ref{bazh_strog}):
 \begin{gather*} R(\lambda,\mu)\,  L^{(1)}_k\, (\lambda) L^{(2)}_k(\mu)=
L^{(2)}_k(\mu)\, L^{(1)}_k(\lambda)\, R(\lambda,\mu),
\end{gather*}
where $L^{(1)}_k(\lambda)= L_k(\lambda)\otimes\mathbb{I}$,
$L^{(2)}_k(\mu)=\mathbb{I} \otimes L_k(\mu)$. In fact the formulas
for $L$-operators and $R$-matrix given in this contribution are
close to formulas from the paper by Tarasov \cite{Tarasov}.
Original formulas given in \cite{BS} and \cite{Kor} are a bit dif\/ferent (but
equivalent). The monodromy matrix for the BSQC with $n$ particles
is def\/ined as
\begin{gather}\label{mm}
{T}_n(\lambda)=L_1(\lambda)L_2(\lambda)\cdots L_n(\lambda)= \left(
\begin{array}{cc}
A_n(\lambda)& B_n(\lambda)\\[2mm]
C_n(\lambda)& D_n(\lambda) \end{array} \right)
\end{gather}
and satisf\/ies the same intertwining relation
\begin{gather}
R(\lambda,\mu) \, {T}^{(1)}_n(\lambda)\, {T}^{(2)}_n(\mu)=
 {T}^{(2)}_n(\mu)\, {T}^{(1)}_n(\lambda)\, R(\lambda,\mu).
\label{rtt}
\end{gather}

The intertwining relation (\ref{rtt}) gives
$[A_n(\lambda),A_n(\nu)]=0$. Therefore $A_n(\lambda)$ is the
generating function for the commuting set of operators
${\boldsymbol H}_1,\ldots,{\boldsymbol H}_n$:
\[
A_n(\lambda)=1+\lambda {\boldsymbol H}_1+\lambda^2 {\boldsymbol
H}_2+ \cdots+\lambda^n {\boldsymbol H}_n.
\]
We interpret these operators ${\boldsymbol
H}_1,\ldots,{\boldsymbol H}_{n}$ as Hamiltonians of the open BSQC.
The simplest Hamiltonians are
\begin{gather*}
{\boldsymbol H}_1=\sum_{k=1}^n  \varkappa_k {\boldsymbol v}_k
+\sum_{1\le l<k\le n}\!\! {\boldsymbol
u}_l^{-1}(a_l-b_l{\boldsymbol v}_l)\prod_{s=l+1}^{k-1}\frac{b_s
d_s}{\varkappa_s}{\boldsymbol v}_s\cdot {\boldsymbol u}_k
(c_k-d_k{\boldsymbol v}_k), \qquad {\boldsymbol H}_n=\prod_{k=1}^n
\varkappa_k {\boldsymbol v}_k.
\end{gather*}
At $b_k=0$ and $c_k=0$, the BSQC model reduces to Relativistic
Toda Chain (RTC) at root of unity. The corresponding $L$-operators
are
\begin{gather*}
L_k^{\rm RTC}(\lambda)=\left(
\begin{array}{cc}
1+\lambda \varkappa_k {\boldsymbol v}_k &  \lambda a_k {\boldsymbol u}_k^{-1} \\
- d_k {\boldsymbol u}_k {\boldsymbol v}_k  & 0
\end{array}
\right),\qquad k=1,2,\ldots,n.
\end{gather*}
As in BSQC model $A_n^{\rm RTC}(\lambda)$ is the generating
function for the commuting set of operators ${\boldsymbol
H}_1^{\rm RTC},\ldots,{\boldsymbol H}_n^{\rm RTC}$. The simplest
Hamiltonians for RTC are
\[
{\boldsymbol H}_1^{\rm RTC}=\sum_{k=1}^n  \varkappa_k {\boldsymbol
v}_k -\sum_{1\le l \le n-1} a_l d_{l+1} {\boldsymbol
u}_l^{-1}{\boldsymbol u}_{l+1}{\boldsymbol v}_{l+1}, \qquad
{\boldsymbol H}_n^{\rm RTC}=\prod_{k=1}^n \varkappa_k {\boldsymbol
v}_k.
\]
Note that these operators ${\boldsymbol H}_1^{\rm
RTC},\ldots,{\boldsymbol H}_{n}^{\rm RTC}$ are the Hamiltonians of
the RTC with open boun\-dary condition and association with the
standard operators of momenta ${\boldsymbol p}_k$ and positions
${\boldsymbol q}_k$ roughly speaking (up to constants) is
${\boldsymbol v}_k=\exp {\boldsymbol p}_k$ and ${\boldsymbol
u}_k=\exp {\boldsymbol q}_k$. Then ${\boldsymbol H}_n^{\rm RTC}$
is the exponent of the total momentum of RTC and ${\boldsymbol
H}_1^{\rm RTC}$ is the Hamiltonian of relativistic analogue of
usual Toda chain.

\section{Eigenvalues and associated amplitudes}

In this section we give a procedure of obtaining the eigenvalues
for open $n$-particle BSQC Hamiltonians ${\boldsymbol H}_k$,
$k=1,2,\ldots,n$, or equivalently for $A_n(\lambda)$.

In the case of $a_k=b_k=c_k=d_k=0$, $k=1,2,\ldots,n$, we have
$A_n(\lambda)=\prod\limits_{k=1}^n (1+\varkappa_k{\boldsymbol
v}_k)$. We interpret the corresponding Hamiltonians as free
(without interaction between particles) Hamiltonians. Due to
(\ref{uvact}) the standard basis vectors
$|\gamma_1,\gamma_2,\ldots,\gamma_n\rangle =
|\gamma_1\rangle_1\otimes |\gamma_2\rangle_2\otimes \cdots \otimes
|\gamma_n\rangle_n\in {\cal V}$ are eigenvectors of
$A_n(\lambda)$:
\[
A_n(\lambda) |\gamma_1,\gamma_2,\ldots,\gamma_n\rangle =
\prod_{k=1}^n (1+\varkappa_k\omega^{\gamma_k})
|\gamma_1,\gamma_2,\ldots,\gamma_n\rangle.
\]

We claim that in the general case the spectrum of $A_n(\lambda)$
has the form as in the case of non-interacting particles but with
modif\/ied amplitudes $\varkappa_{n,k}$:
\[
A_n(\lambda)|\gamma_1,\gamma_2,\ldots,\gamma_n\rangle
=\prod_{k=1}^n\left(1+\varkappa_{n,k}\omega^{\gamma_{k}}\lambda\right)
|\gamma_1,\gamma_2,\ldots,\gamma_n\rangle,\qquad \gamma_{k}\in
\mathbb{Z}_N.
\]
The corresponding eigenvectors
$|\gamma_1,\gamma_2,\ldots,\gamma_n\rangle$ are not standard basis
vectors of course. To obtain their coordinates in the standard
basis we will use an iterative procedure as was promised in the
introduction. We start from the eigenvectors of open $1$-particle
BSQC. Then we construct eigenvectors of open $2$-particle BSQC by
addition in an appropriate way one more particle. And so on. In
parallel with this procedure we have an iterative procedure of
obtaining the amplitudes:
\[
(\varkappa_{11}:=\varkappa_1)\stackrel{+2^{\rm nd} {\rm \
particle}} \longrightarrow(\varkappa_{21},\varkappa_{22})
\stackrel{+3^{\rm rd} {\rm \ particle}}\longrightarrow \cdots
\stackrel{+n^{\rm th} {\rm \ particle}}\longrightarrow
 (\varkappa_{n1},\ldots,\varkappa_{nn}).
\]
Now we will describe the procedure how to f\/ind these amplitudes
$\varkappa_{m,s}$, $m=1,2,\ldots,n$, $s=1,2,\ldots,m$. We will
need the variables
\begin{gather}\label{xxx}
x^{m,s}_{m',s'}=\frac{\varkappa_{m,s}}{\varkappa_{m',s'}},\qquad
x_m=\frac{c_m}{d_m},\qquad x_{m,s}=\frac{c_m\varkappa_m}{d_m
\varkappa_{m,s}},\qquad \tilde x_{m,s}=\frac{b_m
\varkappa_{m,s}}{a_m\varkappa_m},
\end{gather}
and variables $y^{m,s}_{m',s'}$, $y_m$, $y_{m,s}$, $\tilde
y_{m,s}$. The latter are def\/ined (up to a root of $1$, which
will be f\/ixed later) by condition that points
$p^{m,s}_{m',s'}=(x^{m,s}_{m',s'},y^{m,s}_{m',s'})$, $p_m=(x_m,
y_m)$, $p_{m,s}=(x_{m,s}, y_{m,s})$, $\tilde p=(\tilde
x_{m,s},\tilde y_{m,s})$ belong to Fermat curve $x^N+y^N=1$.
First, we def\/ine $\varkappa_{1,1}:=\varkappa_1$. If we
constructed all the above variables for $m-1$ particles, we
def\/ine the variables $\varkappa_{m,1},\varkappa_{m,2},\ldots,
\varkappa_{m,m}$ by the equations
\begin{gather}\label{rel_m}
\varkappa_{m,1}\varkappa_{m,2}\cdots
\varkappa_{m,m}=\varkappa_{m-1,1}\varkappa_{m-1,2} \cdots
\varkappa_{m-1,m-1} \varkappa_m,
\\
\label{rel_other} \frac{\varkappa_m }{a_{m-1}d_m} \frac{y_{m-1}}{
y_m y_{m-1,l}\tilde y_{m-1,l}} \prod_{s\ne l}
\frac{y^{m-1,s}_{m-1,l}}{y^{m-1,l}_{m-1,s}}
\frac{\prod\limits_{k=1}^m y^{m-1,l}_{m,k}}
{\prod\limits_{s=1}^{m-2} y^{m-2,s}_{m-1,l}}=1, \qquad
l=1,2,\ldots,m-1.
\end{gather}
We would like to mention that this iterative procedure has a similarity to iterative procedures
in \cite{PS,GPS}.
To solve these equations we f\/irst take $N$-th power of them. It
gives us system of linear equations
\begin{gather}\label{rel_mNi}
\varkappa_{m,1}^N \varkappa_{m,2}^N \cdots \varkappa_{m,m}^N=
\varkappa_{m-1,1}^N \varkappa_{m-1,2}^N\cdots
\varkappa_{m-1,m-1}^N \varkappa_m^N,
\\
\label{rel_otherNi} \frac{\varkappa_{m-1,l}^N }{a_{m-1}^N d_m^N}
\frac{y_{m-1}^N}{ y_m^N y_{m-1,l}^N\tilde y_{m-1,l}^N}
\frac{\prod\limits_{k=1}^m
\big(1-\varkappa_{m,k}^N/\varkappa_{m-1,l}^N\big)}
{\prod\limits_{s=1}^{m-2}
\big(1-\varkappa_{m-2,s}^N/\varkappa_{m-1,l}^N\big)}=1, \qquad
l=1,2,\ldots,m-1.
\end{gather}
with respect to elementary symmetric polynomials in variables
$\{\varkappa_{m,1}^N,\ldots,\varkappa_{m,m}^N\}$. Solving equation
with coef\/f\/icients being the values of the mentioned symmetric
polynomials we obtain the values of
$\{\varkappa_{m,1}^N,\ldots,\varkappa_{m,m}^N\}$. The variables
$\{\varkappa_{m,1},\ldots,\varkappa_{m,m}\}$ can be found up to
$N$-th roots of~$1$. We f\/ix their phases in a way to satisfy
(\ref{rel_m}) and (\ref{rel_other}).

To compare these formulas with the formulas for eigenvalues
proposed by Tarasov \cite{Tarasov} we consider polynomials ${\cal
A}_m(\lambda^N)$ with zeroes $\epsilon/\varkappa_{m,s}^N$, $s=1,$
$2,\ldots,$ $m,$ where $\epsilon=(-1)^N$:
\begin{gather}\label{zerA}
{\cal A}_m\big(\lambda^N\big)=
\prod_{s=1}^{m}\big(1-\epsilon\,\varkappa_{m,s}^N \lambda^N\big),
\quad m\ge 2;\qquad\! {\cal
A}_1\big(\lambda^N\big)=1-\epsilon\varkappa_1^N \lambda^N;
\qquad\! {\cal A}_0\big(\lambda^N\big)=1.\!\!
\end{gather}
Then the relations (\ref{rel_mNi}) and (\ref{rel_otherNi}) can be
rewritten compactly as recursion relation for ${\cal
A}_m(\lambda^N)$, $m\ge 2$:
\begin{gather} {\cal A}_m\big(\lambda^N\big)=\left(\big(1-\epsilon\varkappa_{m}^N\lambda^N\big)+
\frac{c_m^N-d_m^N}{c_{m-1}^N-d_{m-1}^N} \left(\frac{b_{m-1}^N
d_{m-1}^N}{\varkappa_{m-1}^N}- \epsilon \lambda^N a_{m-1}^N
c_{m-1}^N\right)\right)
{\cal A}_{m-1}\big(\lambda^N\big) \nonumber\\
\phantom{{\cal
A}_m\big(\lambda^N\big)=}{}+\frac{c_{m}^N\!-\!d_{m}^N}{c_{m-1}^N\!-\!d_{m-1}^N}
 \frac{\big(b_{m-1}^N \!-\!\epsilon \lambda^N a_{m-1}^N \varkappa_{m-1}^N \big)\!\big(\epsilon \lambda^N c_{m-1}^N
\varkappa_{m-1}^N\! - \!d_{m-1}^N\big)}{\varkappa_{m-1}^N } {\cal
A}_{m-2}\big(\lambda^N\big).\!\!\!\! \label{relAm} \end{gather}
Indeed, the relation (\ref{rel_mNi}) follows from the relation for
coef\/f\/icients in (\ref{relAm}) at $(\lambda^N)^m$. If we f\/ix
sequentially $\lambda^N=\epsilon/ \varkappa_{m-1,l}^N$,
$l=1,$$2,\ldots,$$m-1$, (that is by the zeroes of ${\cal
A}_{m-1}(\lambda^N)$) we obtain the relations (\ref{rel_otherNi}).

This recursion relation for ${\cal A}_m(\lambda^N)$ can be
obtained by means of averaged $L$-operators \cite{Tarasov}. Using
\[
{\cal L}_m(\lambda^N)=\left( \begin{array}{cc}
1-\epsilon \varkappa_m^N \lambda^N & -\epsilon\lambda^N\big(a_m^N-b_m^N\big)\\[2mm]
c_m^N-d_m^N & b_m^N d_m^N/\varkappa_m^N-\epsilon \lambda^N a_m^N
c_m^N
\end{array} \right)
\]
we def\/ine polynomials ${\cal A}_m(\lambda^N)$, ${\cal
B}_m(\lambda^N)$, ${\cal C}_m(\lambda^N)$ and ${\cal
D}_m(\lambda^N)$ by
\begin{gather}\label{calABCD}
\left(\begin{array}{cc}
{\cal A}_m\big(\lambda^N\big) &  {\cal B}_m\big(\lambda^N\big)\\[2mm]
{\cal C}_m\big(\lambda^N\big) &  {\cal D}_m\big(\lambda^N\big)
\end{array}\right)={\cal L}_1\big(\lambda^N\big) {\cal L}_2\big(\lambda^N\big) \cdots {\cal L}_m\big(\lambda^N\big) .
\end{gather}
In particular, we have
\begin{gather}
{\cal A}_m\big(\lambda^N\big)= \big(1-\epsilon \varkappa_m^N
\lambda^N \big) {\cal A}_{m-1}\big(\lambda^N\big)
+\big(c_m^N-d_m^N\big) {\cal B}_{m-1}\big(\lambda^N\big),\nonumber
\\
{\cal B}_m\big(\lambda^N\big)=
-\epsilon\lambda^N\big(a_m^N-b_m^N\big) {\cal
A}_{m-1}\big(\lambda^N\big) +\big(b_m^N
d_m^N/\varkappa_m^N-\epsilon \lambda^N a_m^N c_m^N\big) {\cal
B}_{m-1}\big(\lambda^N\big).\label{AAB}
\end{gather}
Excluding ${\cal B}_{m-1}(\lambda^N)$ from these two relations we
get
\begin{gather*}
{\cal B}_m\big(\lambda^N\big)= \frac{b_m^N
d_m^N/\varkappa_m^N-\epsilon \lambda^N a_m^N c_m^N}{c_m^N-d_m^N}
{\cal A}_m\big(\lambda^N\big) -\frac{\det {\cal
L}_m\big(\lambda^N\big)}{c_m^N-d_m^N} {\cal
A}_{m-1}\big(\lambda^N\big).
\end{gather*}
Substituting  the right-hand side of this equation with $m$
replaced by $m-1$ instead of ${\cal B}_{m-1}(\lambda^N)$
in~(\ref{AAB}) we get (\ref{relAm}). Therefore two formulas
(\ref{relAm}) and (\ref{calABCD}) for ${\cal A}_m(\lambda^N)$ are
equivalent. Summarizing, in order to f\/ind amplitudes
$\varkappa_{m,s}$, $s=1,\ldots,m$, for some $m$, we have to f\/ind
${\cal A}_m(\lambda^N)$ using (\ref{relAm}) or (\ref{calABCD}).
Then solving equation ${\cal A}_m(\lambda^N)=0$ of $m$th degree
with respect to $\lambda^N$ and taking into account (\ref{zerA})
we can f\/ind $\varkappa^N_{m,s}$, $s=1,\ldots,m$. This gives us
the set $\varkappa_{m,s}$ up to $N$th roots of $1$. At last step,
we have to f\/ix their values in a way to satisfy (\ref{rel_m})
and (\ref{rel_other}).

It seems to the author that the equation ${\cal A}_m(\lambda^N)=0$
can not be solved explicitly in the case of generic parameters. In
the next section, the solution for the homogeneous RTC is given
explicitly. The author does not know other interesting special
cases of parameters which admit explicit solution for the spectrum
of $A_m(\lambda)$. As shown in \cite{GIPS}, it is possible to give
an explicit solution for the spectrum of $B_m(\lambda)$ in the
homogeneous case of $m$-particle Bazhanov--Stroganov quantum
chain.

\section{Amplitudes for the homogeneous Relativistic Toda Chain}

In this section we sketch the method described in \cite{PS} of
obtaining the amplitudes for the homogeneous RTC: $a_k=a$,
$b_k=0$, $c_k=0$, $d_k=d$, $\varkappa_k=\varkappa$. In this case
the amplitudes $\varkappa_{m,s}$, $s=1,\ldots,m$, can be expressed
in terms of solutions of some quadratic equation. Since
\begin{gather}\label{cLRTC}
{\cal L}^{\rm RTC}_k\big(\lambda^N\big)={\cal L}^{\rm
RTC}(\lambda^N)=\left( \begin{array}{cc}
1-\epsilon \varkappa^N \lambda^N & -\epsilon a^N \lambda^N \\[2mm]
-d^N & 0
\end{array} \right),
\end{gather}
we obtain
\begin{gather*}
\left(\begin{array}{cc}
{\cal A}_m\big(\lambda^N\big) &  {\cal B}_m\big(\lambda^N\big)\\[2mm]
{\cal C}_m\big(\lambda^N\big) &  {\cal D}_m\big(\lambda^N\big)
\end{array}\right)=\left({\cal L}^{\rm RTC}\big(\lambda^N\big)\right)^m .
\end{gather*}
Applying the fact that $2\times 2$ matrix $\boldsymbol M$ with
eigenvalues $\mu_+$ and $\mu_-$ satisf\/ies
\[
{\boldsymbol M}^m=\frac{\mu_+^m-\mu_-^m}{\mu_+-\mu_-}{\boldsymbol
M} -\frac{\mu_+^m \mu_--\mu_-^m \mu_+}{\mu_+-\mu_-}{\boldsymbol 1}
\]
for matrix ${\cal L}^{\rm RTC}(\lambda^N)$ we obtain
\begin{gather}\label{Apredv}
{\cal A}_m\big(\lambda^N\big)=\big(1-\epsilon \varkappa^N
\lambda^N \big) \frac{x_+^m-x_-^m}{x_+-x_-}-\frac{x_+^m x_--x_-^m
x_+}{x_+-x_-},
\end{gather}
where $x_+(\lambda^N)$ and $x_-(\lambda^N)$ are eigenvalues of
${\cal L}(\lambda^N)$. These eigenvalues are roots of
characteristic polynomial $x^2-\tau(\lambda^N) x +
\delta(\lambda^N)=0$:
\begin{gather*}
x_{\pm}=\frac{1}{2}\left(\tau\pm \sqrt{\tau^2-4\delta}\right),
\end{gather*}
where, using (\ref{cLRTC}),
\begin{gather}\label{tau}
\tau\big(\lambda^N\big)={\rm tr}\, {\cal L}^{\rm
RTC}\big(\lambda^N\big)=x_1+x_2= 1 - \epsilon \varkappa^N
\lambda^N ,
\\ \label{delta}
\delta\big(\lambda^N\big)= \det {\cal L}^{\rm
RTC}\big(\lambda^N\big) =x_1 x_2= -\epsilon a^Nd^N \lambda^N .
\end{gather}
Taking into account (\ref{tau}) we rewrite (\ref{Apredv}) as
\begin{gather*}
{\cal A}_m\big(\lambda^N\big)=\frac{x_+^{m+1}-x_-^{m+1}}{x_+-x_-}.
\end{gather*}

Introducing the variable $\phi$ by $x_+/x_-=e^{{\rm i}\phi}$ we
f\/ind that roots of ${\cal A}_m$ correspond to roots $\phi_{m,s}$
of $e^{{\rm i}(m+1)\phi}=1$ (without $\phi=0$) that is
\begin{gather}
\label{zerphi} \phi_{m,s}=2\pi s/(m+1), \qquad s=1,2,\ldots,m.
\end{gather}
Now we need to f\/ind an explicit relation between $\lambda^N$ and
$\phi$. We have
\[
\tau+\sqrt{\tau^2-4\delta}=e^{{\rm i}\phi} \left(\tau-
\sqrt{\tau^2-4\delta}\right).
\]
Therefore
\begin{gather}
\label{tdphi} \tau^2=4 \delta \cos^2 \frac \phi 2.
\end{gather}
 Taking into account (\ref{tau}) and (\ref{delta}) we
consider (\ref{tdphi}) as quadratic equation with respect to
$\lambda^N$:
\begin{gather*}
\lambda^{2N} \varkappa^{2N} + 2\epsilon \lambda^N \big(a^N d^N +
a^N d^N\cos\phi -\varkappa^N\big) +1=0.
\end{gather*}
The solution $\lambda^N(\phi)$ of this equation describes the
relation between the variables $\lambda^N$ and $\phi$. Therefore
we can translate the zeroes (\ref{zerphi}) of ${\cal
A}_m(\lambda^N(\phi))$ in terms of variable $\phi$ to zeroes
$\lambda^N(\phi_{m,s})$ in terms of $\lambda^N$. Finally, taking
into account (\ref{zerA}) we f\/ind
\begin{gather*}
 \varkappa_{m,s}^N=\epsilon/ \lambda^N(\phi_{m,s}),\qquad s=1,2,\ldots,m.
\end{gather*}

\section{Eigenvectors and eigenvalues}

In order to give explicit formulas for the eigenvectors of
$A_n(\lambda)$ we remind the def\/inition (see for example
\cite{BB}) of $w_p(s)$ which is an analogue of $\Gamma_q$-function
at root of unity. For any point $p=(x,y)$ of Fermat curve
$x^N+y^N=1$, we def\/ine $w_p(s)$, $s\in \mathbb{Z}_N$, by
\begin{gather}\label{defw}
\frac{w_p(s)}{w_p(s-1)}=\frac{y}{1-x \omega^s},\qquad w_p(0)=1.
\end{gather}
The function $w_p(s)$ is cyclic: $w_p(s+N)=w_p(s)$.

We will use the notation $|\boldsymbol{\gamma}_n\rangle \in{\cal
V}_1 \otimes\cdots\otimes{\cal V}_n$ for eigenvectors of the
operator $A_n(\lambda)$ of the BSQC with $n$ particles. These
eigenvectors are labeled by $n$ parameters $\gamma_{n,s}\in
\mathbb{Z}_N$, $s=1,2,\ldots,n$, collected into a vector
\begin{gather*}
\boldsymbol{\gamma}_n=(\gamma_{n,1},\ldots,\gamma_{n,n})\in(\mathbb{Z}_N)^n.
\end{gather*}

The following theorem gives a procedure of obtaining the
eigenvectors $|\boldsymbol{\gamma}_n\rangle$ of $A_n(\lambda)$
from the eigenvectors $|\boldsymbol{\gamma}_{n-1}\rangle\in{\cal
V}_1 \otimes\cdots\otimes{\cal V}_{n-1}$ of $A_{n-1}(\lambda)$ and
basis vectors $|\gamma_n\rangle_n\in{\cal V}_{n}$. To f\/ind the
formula for $|\boldsymbol{\gamma}_{n-1}\rangle$ we can use the
same theorem and so on. At the last step we need the eigenvectors
of $1$-particle quantum chain.

From (\ref{bazh_strog}) and (\ref{mm}), it is easy to see that  the vectors
$|{\gamma_{1,1}}\rangle_1\in {\cal V}_1$, $\gamma_{1,1}\in
\mathbb{Z}_N$, are eigenvectors for $A_1(\lambda)$:
\[
A_1(\lambda) |{\gamma_{1,1}}\rangle_1 =
(1+\varkappa_{1,1}\omega^{\gamma_{1,1}})\,
|{\gamma_{1,1}}\rangle_1,
\]
where $\varkappa_{1,1}=\varkappa_1$.

In what follows, the vector $\boldsymbol{\gamma}_n^{\pm k}$ means
the vector $\boldsymbol{\gamma}_n$ in which $\gamma_{n,k}$ is
replaced by $\gamma_{n,k}\pm 1$.

\begin{theorem}
The vector
$|\boldsymbol{\gamma}_n\rangle=|\gamma_{n1},\ldots,\gamma_{nn}\rangle$
\begin{gather}\label{formvect}
|\boldsymbol{\gamma}_n\rangle=\sum_{\boldsymbol{\gamma}_{n-1}\in
(\mathbb{Z}_N)^{n-1}}
 Q(\boldsymbol{\gamma}_{n-1}|\boldsymbol{\gamma}_{n}) |\boldsymbol{\gamma}_{n-1}\rangle\otimes
|\sigma_n\rangle_n
\end{gather}
satisfies
\begin{gather}\label{Alm}
A_n(\lambda)|\boldsymbol{\gamma}_n\rangle
=\prod_{k=1}^n\left(1+\varkappa_{n,k}\omega^{\gamma_{n,k}}\lambda\right)
|\boldsymbol{\gamma}_n\rangle
=\prod_{k=1}^n\left(1-\lambda/\lambda_{n,k}\right)
|\boldsymbol{\gamma}_n\rangle,
\\
\label{Blmk}
B_n\left(\lambda_{n,k}\right)|\boldsymbol{\gamma}_n\rangle=
\frac{a_n \lambda_{n,k}}{ y_n}
\big(1-x_{n,k}\omega^{-\gamma_{n,k}-1}\big)\big(1-\tilde x_{n,k}
\omega^{\gamma_{n,k}}\big) \left(\prod_{s=1}^{n-1}
y^{n-1,s}_{n,k}\right)|\boldsymbol{\gamma}_n^{+k}\rangle,
\\
B_n(\lambda)|\boldsymbol{\gamma}_n\rangle=\lambda\frac{a_n}{ y_n}
\sum_{k=1}^n\left(\prod_{s\ne k}
\frac{\lambda-\lambda_{n,s}}{\lambda_{n,k}-\lambda_{n,s}}\right)\nonumber\\
\label{Blm}
\phantom{B_n(\lambda)|\boldsymbol{\gamma}_n\rangle=}{}\times
\big(1-x_{n,k}\omega^{-\gamma_{n,k}-1}\big)\big(1-\tilde x_{n,k}
\omega^{\gamma_{n,k}}\big) \left(\prod_{l=1}^{n-1}
y^{n-1,l}_{n,k}\right)|\boldsymbol{\gamma}_n^{+k}\rangle,\qquad
n>1,
\\
\label{Blm1} B_1(\lambda)|\boldsymbol{\gamma}_1\rangle=\lambda
a_1\big(1-\tilde x_{1,1} \omega^{\gamma_{1,1}}\big)
 |\boldsymbol{\gamma}_1^{+1}\rangle,
\end{gather}
if
$|\boldsymbol{\gamma}_{n-1}\rangle=|\gamma_{n-1,1},\ldots,\gamma_{n-1,n-1}\rangle\in{\cal
V}_1 \otimes\cdots\otimes{\cal V}_{n-1}$ satisfies the same
relations with $n$ replaced by $n-1$. In the above formulas we
used
\begin{gather}
Q(\boldsymbol{\gamma}_{n-1}|\boldsymbol{\gamma}_{n})=\frac{\omega^{\gamma_{n-1,1}+\cdots+\gamma_{n-1,n-1}}
\prod\limits_{l=1}^{n-1} \prod\limits_{k=1}^n
w_{p^{n-1,l}_{n,k}}(\gamma_{n-1,l}-\gamma_{n,k})}
{w_{p_n}(-\sigma_n-1) \prod\limits_{j,l=1\atop (j\ne l)}^{n-1}
w_{p^{n-1,j}_{n-1,l}}(\gamma_{n-1,j}-\gamma_{n-1,l})}\cdot
\frac{\prod\limits_{l=1}^{n-1} w_{p_{n-1,l}}(-\gamma_{n-1,l}-1)}
{\prod\limits_{l=1}^{n-1} w_{\tilde p_{n-1,l}}(\gamma_{n-1,l}-1)},
\nonumber\\
Q(\boldsymbol{\gamma}_1|\boldsymbol{\gamma}_2)=\frac{\omega^{\gamma_{1,1}}
w_{p^{1,1}_{2,1}}(\gamma_{1,1}-\gamma_{2,1})w_{p^{1,1}_{2,2}}(\gamma_{1,1}-\gamma_{2,2})
} {w_{p_2}(\gamma_{1,1}-\gamma_{2,1}-\gamma_{2,2}-1) w_{\tilde
p_{1,1}}(\gamma_{1,1}-1)},
\nonumber\\
\lambda_{m,s}=-\omega^{-\gamma_{m,s}}/\varkappa_{m,s}, \qquad
\sigma_n(\boldsymbol{\gamma}_{n-1}, \boldsymbol{\gamma}_n)\equiv
\sigma_n=\sum_{k=1}^n \gamma_{n,k} - \sum_{l=1}^{n-1}
\gamma_{n-1,l}. \label{exprQ}
\end{gather}
\end{theorem}

\begin{proof}
We suppose that the formulas (\ref{Alm}) and (\ref{Blm}) with $n$
replaced by $n-1$ are proved. To prove the action formulas
(\ref{Alm}) and (\ref{Blm}) we use the recurrent relations
\begin{gather}\label{recA}
A_n(\lambda)=A_{n-1}(\lambda)\left(1+\lambda \varkappa_n
{\boldsymbol v}_n\right) +B_{n-1}(\lambda)\,{\boldsymbol u}_n
(c_n- d_n{\boldsymbol v}_n),
\\
\label{recB} B_n(\lambda)=A_{n-1}(\lambda)\,\lambda{\boldsymbol
u}_n^{-1}(a_n-b_n {\boldsymbol v}_n) +B_{n-1}(\lambda)
\left(\lambda a_n c_n+\frac {b_n d_n}{\varkappa_n} {\boldsymbol
v}_n\right)
\end{gather}
which follow from (\ref{mm}).

{\it The action formula for $A_n(\lambda)$}: To prove the action
formula (\ref{Alm}) we act by both sides of~(\ref{recA})
on~(\ref{formvect}) and use the formulas (\ref{Alm}) and
(\ref{Blm}) with $n$ replaced by $n-1$. After shifting in an
appropriate way the variables of summation
$\boldsymbol{\gamma}_{n-1}$ we reduce the problem to
verif\/ication of relation
\begin{gather}
\left( \prod_{k=1}^n\left(1-\lambda/\lambda_{n,k}\right)-
\prod_{l=1}^{n-1}\left(1-\lambda/\lambda_{n-1,l}\right) (1+\lambda
\varkappa_n \omega^{\sigma_n})
\right)Q(\boldsymbol{\gamma}_{n-1}|\boldsymbol{\gamma}_{n})\nonumber\\
\qquad{}=\lambda \frac{a_{n-1}}{ y_{n-1}}
\sum_{l=1}^{n-1}\left(\prod_{s=1\atop s\ne l}^{n-1}
\frac{\lambda-\lambda_{n-1,s}}
{\omega\lambda_{n-1,l}-\lambda_{n-1,s}}\right)
\big(1-x_{n-1,l}\omega^{-\gamma_{n-1,l}}\big)\big(1-\tilde x_{n-1,l} \omega^{\gamma_{n-1,l}-1}\big)\nonumber\\
\qquad\quad{}\times \left(\prod_{s=1}^{n-2}
y^{n-2,s}_{n-1,l}\right) \big(c_n-d_n \omega^{\sigma_n+1}\big)
Q(\boldsymbol{\gamma}_{n-1}^{-l}|\boldsymbol{\gamma}_{n}).\label{idQ}
\end{gather}
Using
\begin{gather*}
\frac{Q(\boldsymbol{\gamma}_{n-1}^{-l}|\boldsymbol{\gamma}_{n})}{Q(\boldsymbol{\gamma}_{n-1}|\boldsymbol{\gamma}_{n})}
\frac{a_{n-1}}{ y_{n-1}}
(1-x_{n-1,l}\omega^{-\gamma_{n-1,l}})(1-\tilde x_{n-1,l}
\omega^{\gamma_{n-1,l}-1})
\left(\prod_{s=1}^{n-2} y^{n-2,s}_{n-1,l}\right)\nonumber\\
\qquad{}{} =\frac{\varkappa_n}{\omega d_n}
\frac{\prod\limits_{k=1}^{n}\left(1-\frac{\lambda_{n,k}}{\lambda_{n-1,l}}\right)}
{1-x_n\omega^{-\sigma_n-1}} \prod_{s\ne
l}\frac{1-\omega\frac{\lambda_{n-1,l}}{\lambda_{n-1,s}}}
{1-\frac{\lambda_{n-1,s}}{\lambda_{n-1,l}}}
\end{gather*}
which follows directly from (\ref{exprQ}) and (\ref{rel_other}),
we rewrite (\ref{idQ}) as
\begin{gather*}
\prod_{k=1}^n\left(1-\lambda/\lambda_{n,k}\right)-
\prod_{l=1}^{n-1}\left(1-\lambda/\lambda_{n-1,l}\right) (1+\lambda
\varkappa_n \omega^{\sigma_n})
\\
\qquad{}=\lambda \sum_{l=1}^{n-1}\left(\prod_{s=1\atop s\ne
l}^{n-1} \frac{\lambda-\lambda_{n-1,s}}
{\lambda_{n-1,l}-\lambda_{n-1,s}}\right) \frac{-\varkappa_n
\omega^{\sigma_n}}{\lambda_{n-1,l}} \frac{\prod\limits_{k=1}^n
\lambda_{n,k}}{\prod\limits_{s=1}^{n-1} \lambda_{n-1,s}}
\prod_{k=1}^n\left(1-\lambda_{n-1,l}/\lambda_{n,k}\right) .
\end{gather*}
Taking into account
\begin{gather}\label{kpnkpn-1}
-\varkappa_n \omega^{\sigma_n} \frac{\prod\limits_{k=1}^n
\lambda_{n,k}}{\prod\limits_{s=1}^{n-1} \lambda_{n-1,s}}=1,
\end{gather}
which follows from (\ref{rel_m}), we obtain f\/inally
\begin{gather*}
\prod_{k=1}^n\left(1-\lambda/\lambda_{n,k}\right)-
\prod_{l=1}^{n-1}\left(1-\lambda/\lambda_{n-1,l}\right) (1+\lambda
\varkappa_n \omega^{\sigma_n})
\\
\qquad{}=\lambda \sum_{l=1}^{n-1}\left(\prod_{s=1\atop s\ne
l}^{n-1} \frac{\lambda-\lambda_{n-1,s}}
{\lambda_{n-1,l}-\lambda_{n-1,s}}\right) \frac{1}{\lambda_{n-1,l}}
\prod_{k=1}^n\left(1-\lambda_{n-1,l}/\lambda_{n,k}\right) .
\end{gather*}
To verify this equality  we note that both sides are polynomials
in $\lambda$ of degree $n-1$ (not $n$ due to~(\ref{kpnkpn-1}))
without free term. Therefore it is suf\/f\/icient to verify this
relation at $n-1$ dif\/ferent values of $\lambda$. Taking these
values to be $\lambda=\lambda_{n-1,l}$, $l=1,2,\ldots,n-1$, we
easily prove the relation. Thus we proved (\ref{Alm}).

{\it The action formula for $B_n(\lambda_{n,k})$}: Next we show
the validity of (\ref{Blmk}). The action formulas
for~$B_1(\lambda)$ and $B_2(\lambda)$ can be verif\/ied in a
direct way. Thus we suppose $n>2$.

Excluding $B_{n-1}(\lambda)$ from (\ref{recA}) and substituting it
into (\ref{recB}) we get
\begin{gather}
{\boldsymbol u}_n (c_n- d_n{\boldsymbol v}_n) B_n(\lambda)=
\left(\lambda a_n c_n+\omega \frac {b_n d_n}{\varkappa_n}
{\boldsymbol v}_n\right)A_n(\lambda)
\nonumber\\
\qquad{} -\omega \lambda a_n d_n {\boldsymbol v}_n \left(1+
\frac{\lambda c_n \varkappa_n}{\omega d_n}\right)
\left(1+\frac{b_n}{\lambda
a_n\varkappa_n}\right)A_{n-1}(\lambda).\label{recAB}
\end{gather}

Let us apply (\ref{recAB}) to $|\boldsymbol{\gamma}_n\rangle$ for
$\lambda=\lambda_{n,k}=-\omega^{-\gamma_{n,k}}/\varkappa_{n,k}$,
i.e.\ at the zeros of eigenvalue of~$A_n(\lambda)$. This gives, by
virtue of the def\/initions (\ref{xxx}) of $x_{m,s}$ and $\tilde
x_{m,s}$:
\begin{gather}
{\boldsymbol u}_n (c_n- d_n{\boldsymbol v}_n) B_n(\lambda_{n,k})
|\boldsymbol{\gamma}_n\rangle= -\lambda_{n,k} \omega a_n d_n
\big(1-x_{n,k}\omega^{-\gamma_{n,k}-1}\big)
\big(1-\tilde x_{n,k} \omega^{\gamma_{n,k}}\big)\nonumber\\
\qquad{}{}\times  \sum_{\boldsymbol{\gamma}_{n-1}\in
(\mathbb{Z}_N)^{n-1}}\!\!\omega^{\sigma_n}
 Q(\boldsymbol{\gamma}_{n-1}|\boldsymbol{\gamma}_n) A_{n-1}(\lambda_{n,k})
 |\boldsymbol{\gamma}_{n-1}\rangle\otimes |\sigma_n\rangle. \label{ABm}
\end{gather}

From (\ref{Alm}) we know how to apply
$A_{n-1}$ to $|\boldsymbol{\gamma}_{n-1}\rangle$:
\begin{gather}\label{Almnk}
A_{n-1}(\lambda_{n,k})|\boldsymbol{\gamma}_{n-1}\rangle =
 \prod_{s=1}^{n-1}\left(1-\frac{\varkappa_{n-1,s}}{\varkappa_{n,k}}\omega^{\gamma_{n,k}-\gamma_{n-1,s}}\right)
 |\boldsymbol{\gamma}_{n-1}\rangle.
\end{gather}
Using (\ref{uvact}) we f\/ind the action of the inverse of the
operator ${\boldsymbol u}_n (c_n-d_n {\boldsymbol v}_n)$ on
$|\sigma_n\rangle_n$\,:
\begin{gather}\label{invop}
({\boldsymbol u}_n (c_n-d_n {\boldsymbol v}_n))^{-1}
|\sigma_n\rangle_n= \big(c_n-d_n \omega^{\sigma _n+1}\big)^{-1}
|\sigma_n+1\rangle_n.
\end{gather}
Taking into account (\ref{exprQ}) and (\ref{defw}) we get
\begin{gather}\label{QQ}
 Q(\boldsymbol{\gamma}_{n-1}|\boldsymbol{\gamma}_n)
 \frac{\omega^{\sigma_n}\prod\limits_{s=1}^{n-1}
 \left(1-\frac{\varkappa_{n-1,s}}{\varkappa_{n,k}}\omega^{\gamma_{n,k}-\gamma_{n-1,s}}\right)}
 {c_n-d_n \omega^{\sigma_n+1}}=
 -\omega^{-1}\,Q(\boldsymbol{\gamma}_{n-1}|\boldsymbol{\gamma}_n^{+k})
 \frac{\prod\limits_{s=1}^{n-1} y^{n-1,s}_{n,k}}{d_n y_n}.
\end{gather}
Finally, using (\ref{Almnk}), (\ref{invop}) and (\ref{QQ}), we
reduce (\ref{ABm}) to (\ref{Blmk}).

{\it The action formula for $B_n(\lambda)$}:
From (\ref{bazh_strog}) and (\ref{mm}) it is easy to f\/ind that
the operator $B_n(\lambda)/\lambda$ is a polynomial in $\lambda$
of $(n-1)$th order. Due to (\ref{Blmk}) we know the action
formulas for $B_n(\lambda)/\lambda$ at the $n$ particular values
of $\lambda$: $\lambda=\lambda_{n,k}$, $k=1,$ $2,\ldots,n$. This
data is enough to reconstruct the action of the polynomial
$B_n(\lambda)$ on $|\boldsymbol{\gamma}_n\rangle$ uniquely.
Lagrange interpolation formula gives
\[
\frac{B_n(\lambda)}{\lambda}|\boldsymbol{\gamma}_n\rangle=\sum_{k=1}^n\left(\prod_{l\ne
k}
\frac{\lambda-\lambda_{n,l}}{\lambda_{n,k}-\lambda_{n,l}}\right)
\frac{B_n(\lambda_{n,k})}{\lambda_{n,k}}|\boldsymbol{\gamma}_n\rangle.
\]
Finally using (\ref{Blmk}) we get (\ref{Blm}). This completes the
proof of the Theorem.
\end{proof}

\section{Discussion}
In this contribution we applied the iterative procedure
of obtaining the eigenvectors for quantum integrable systems
by Kharchev and Lebedev~\cite{KhL}
(which has origin in~\cite{Skl} by Sklyanin)
to open Bazhanov--Stroganov quantum chain.
We plan to extend this result (along the line of the paper~\cite{IS})
to the case of Bazhanov--Stroganov chain with
integrable boundary interaction.

\subsection*{Acknowledgements}

The author would like to acknowledge the organizers of the Sixth
International Conference ``Symmetry in Nonlinear Mathematical
Physics'' (June 20--26, Kyiv) for their nice conference. The
present paper is the written version of the talk delivered by the
author at this conference. The author is thankful to
Professors G.~von~Gehlen, S.~Pakuliak and V.~Shadura for
collaboration in obtaining the results presented in this
contribution. The research presented here is partially supported
by INTAS (grant No.03-51-3350) and by the French--Ukrainian
project ``Dnipro''.

\LastPageEnding

\end{document}